\documentclass[12pt,onecolumn,a4paper]{IEEEtran}
\usepackage{amsmath}
\usepackage{graphicx}
\renewcommand{\baselinestretch}{1.8}

\newcommand{\figuramedia}[3]
{
\begin{figure}
  \centering
 \includegraphics[width=12cm]{#1}
  \caption{#2}\label{#3}
\end{figure}
}

\newcommand{\figura}[3]
{
\begin{figure}
  \centering
 \includegraphics[width=12cm]{#1}
  \caption{#2}\label{#3}
\end{figure}
}
\title{On the Optimization of the IEEE 802.11 DCF: A Cross-Layer Perspective}
\author{\authorblockN{Massimiliano Laddomada and Fabio Mesiti
\thanks{Massimiliano Laddomada is with the Electrical Engineering Dept. of Texas A\&M University-Texarkana, email: \textrm{mladdomada@tamut.edu}.}
\thanks{Fabio Mesiti is with DELEN, Politecnico di Torino, Italy.}}}
\begin{document}
\maketitle

\begin{abstract}
This paper is focused on the problem of optimizing the aggregate
throughput of the Distributed Coordination Function (DCF)
employing the basic access mechanism at the data link layer of
IEEE 802.11 protocols. In order to broaden the applicability of
the proposed analysis, we consider general operating conditions
accounting for both non-saturated and saturated traffic in the
presence of transmission channel errors, as exemplified by the
packet error rate $P_e$.

The main clue of this work stems from the relation that links the
aggregate throughput of the network to the packet rate $\lambda$
of the contending stations. In particular, we show that the
aggregate throughput $S(\lambda)$ presents two clearly distinct
operating regions that depend on the actual value of the packet
rate $\lambda$ with respect to a critical value $\lambda_c$,
theoretically derived in this work.

The behavior of $S(\lambda)$ paves the way to a cross-layer
optimization algorithm, which proved to be effective for
maximizing the aggregate throughput in a variety of network
operating conditions. A nice consequence of the proposed
optimization framework relies on the fact that the aggregate
throughput can be predicted quite accurately with a simple, yet
effective, closed-form expression, which is also derived in the
article.

Finally, theoretical and simulation results are presented
throughout the work in order to unveil, as well as verify, the key
ideas.
\end{abstract}
\section{Introduction}
DCF represents the main access mechanism at the Medium Access
Control (MAC) layer~\cite{standard_DCF_MAC} of the IEEE802.11
series of standards, and it is based on Carrier Sense Multiple
Access with Collision Avoidance (CSMA/CA).
%
%%%%%%%%% LITERATURE REVIEW     %%%%%%%%%
%%%%%%%%%%%%%%%%%%%%%%%%%%%%%%%%%%%%%%%%%%%%%%%%%%%%%%%%%%%%%%%%%%%%%%%%%%%%

Many papers, following the seminal work by Bianchi~\cite{Bianchi},
have addressed the problem of modeling, as well as optimizing, the
DCF in a variety of traffic load models and transmission channel
conditions.

Let us provide a survey of the recent literature related to the
problem addressed in this paper. Papers
\cite{Daneshgaran_sat}-\cite{HCheolLee} model the influence of
real channel conditions on the throughput of the DCF operating in
saturated traffic conditions. Paper \cite{Daneshgaran_sat}
investigates the saturation throughput of IEEE 802.11 in presence
of non ideal transmission channel and capture effects.

The behavior of the DCF of IEEE 802.11 WLANs in unsaturated
traffic conditions has been analyzed in
\cite{Liaw}-\cite{Daneshgaran_unsat}, whereby the authors proposed
various bi-dimensional Markov models for unsaturated traffic
conditions, extending the basic bi-dimensional model proposed by
Bianchi~\cite{Bianchi}. In \cite{Qiao}, the authors look at the
impact of channel induced errors and of the received
Signal-to-Noise ratio (SNR) on the achievable throughput in a
system with rate adaptation, whereby the transmission rate of the
terminal is modified depending on either direct, or indirect
measurements of the link quality.

The effect of the contention window size on the performance of the
DCF has been investigated in \cite{cali}-\cite{choudhury} upon
assuming a variety of transmission scenarios.
In~\cite{cali}-\cite{Khalaj2}, the authors addressed the
optimization of the contention window in such a way as to optimize
the saturated throughput of the DCF.
%
%%%%%%%%%%%%%%%%%%%%%%%%%%%%%%%%%%%%%%%%%%%%%%%%%%%%%%%%%%%%%%%%%%%%%%%%%%%%
%%%%%%%%%%%%%%%%%%%%%%%%%%%%%%%%%%%%%%%%%%%%%%%%%%%%%%%%%%%%%%%%%%%%%%%%%%%%

As a starting point for the derivations which follow, we adopt the
bi-dimensional model proposed in a companion
paper~\cite{daneshgaran} (see also \cite{Daneshgaran_unsat}).
Briefly, in~\cite{daneshgaran} the authors extended the saturated
Bianchi's model by introducing a new idle state, not present in
the original Bianchi's model, accounting for the case in which the
station queue is empty. The Markov chain of the proposed
contention model was solved for obtaining the stationary
probabilities, along with the probability $\tau$ that a station
starts transmitting in a randomly chosen time slot.

In this paper the focus is different in many respects. Firstly, we
investigate the behavior of the aggregate throughput as a function
of the traffic load $\lambda$ of the contending stations, and note
that the aggregate throughput $S(\lambda)$ presents two distinct
operating regions identified, respectively, by Below Link Capacity
(BLC) region and Link Capacity (LC) region. Derived in closed-form
in this work, and identified by $S_m$ throughout the paper, the
link capacity of the considered scenario corresponds to the
maximum throughput that the network can achieve when the
contending stations transmit with a proper set of network
parameters. We show that the network operates in the BLC region
when the actual value of the packet rate $\lambda$ is less than a
critical value $\lambda_c$, theoretically derived in this work.

The second part of this article is focused on the optimization of
the DCF throughput. We propose a cross-layer algorithm whose main
aim is to allow the network to operate as close as possible to the
link capacity $S_m$. The proposed optimization algorithm relies on
a number of insights derived by the behavior of the aggregate
throughput $S(\lambda)$ as a function of the traffic load
$\lambda$, and aims at choosing either an appropriate value of the
minimum contention window $W_0$, or a proper size of the
transmitted packets depending on the network operating region.

Finally, we derive a simple model of the optimized throughput,
which is useful for predicting the aggregate throughput without
resorting to simulation.

The rest of the paper is organized as follows.
Section~\ref{Section_Problem_Formulation} briefly presents the
Markov model at the very basis of the proposed optimization
framework, whereas Section~\ref{Section_Throughput_Analysis}
investigates the behavior of the aggregate throughput as a
function of the traffic load $\lambda$. The proposed optimization
algorithm is discussed in
section~\ref{Section_throughput_Optimization}.
Section~\ref{Section_simulation_results} presents simulation
results of the proposed technique applied to a sample network
scenario, while Section~\ref{Section_Conclusions} draws the
conclusions.
\section{Markov Modeling}
\label{Section_Problem_Formulation}
%
%%
%\figuramedia{unsat__2WAY__Wopt__letter_R.eps}{Markov chain for the
%contention model accounting for unsaturated traffic conditions.
%The model considers $m$ backoff stages. $P_{eq}$ is the
%probability of failed transmission, accounting for both collisions
%and channel errors affecting the transmitted packets.}{fig.chain}
%%
The bi-dimensional Markov Process of the contention model proposed
in \cite{daneshgaran}, and shown in Fig.~\ref{fig.chain} for
completeness, governs the behavior of each contending station
through a series of states indexed by the pair $(i,k),~\forall
i\in [0,m], k\in [0,W_i-1],$ whereby $i$ identifies the backoff
stage. On the other hand, the index $k$, which belongs to the set
$[0,W_i-1]$, identifies the backoff counter. By this setup, the
size of the $i$th contention window is $W_i=2^i W_0,~\forall i\in
[1,m],$ while $W_0$ is the minimum size of the contention window.

An idle state, identified by $I$ in Fig.~\ref{fig.chain}, is
introduced in order to account for the scenario in which after a
successful transmission there are no packets to be transmitted, as
well as for the situation in which the packet queue is empty and
the station is waiting for a new packet arrival.

The proposed Markov model accounts for packet errors due to
imperfect channel conditions, by defining an equivalent
probability of failed transmission, identified by $P_{eq}$, which
considers the need for a new contention due either to packet
collisions ($P_{col}$) or to channel errors ($P_e$) on the
transmitted packets, i.e.,
\begin{equation}%\small
\label{eq.equ}
P_{eq}=P_{col}+P_e-P_e\cdot  P_{col}
\end{equation}
It is assumed that at each transmission attempt each station
encounters a constant and independent probability of failed
transmission, independently from the number of retransmissions
already suffered.

The stationary solution of the Markov Model in
Fig.~\ref{fig.chain} is employed to compute $\tau$, the
probability that a station starts a transmission in a randomly
chosen time slot. First of all, observe that a packet transmission
takes place when a station goes in one of the states $b_{i,0},
\forall i\in\{0,m\}$. Therefore, the probability $\tau$ can be
evaluated by adding up the probabilities $b_{i,0}$ over the set
$i\in \{0,\ldots,m\}$. Upon imposing the normalization
condition\footnote{The probabilities $b_{i,k}$ must add up to $1$
$\forall k\in \{1,\ldots,W_i-1\}$, and $\forall i\in\{0,m\}$.} on
the Markov model in Fig.~\ref{fig.chain}, and expressing the
probabilities $b_{i,k}, \forall k\in \{1,\ldots,W_i-1\},$ as a
function of the probabilities $b_{i,0}$, the probability $\tau$
can be rewritten as follows:
\begin{equation}%\small
\label{eq.tau}
\begin{array}{lll}
  \tau &= &\frac{2(1-2P_{eq})q}{q[(W_0+1)(1-2P_{eq}) + W_0 P_{eq}(1-(2P_{eq})^m)] + 2(1-q)(1-P_{eq})(1-2P_{eq})}
\end{array}
\end{equation}
As in Bianchi's work \cite{Bianchi}, we assume that: 1) the
probability $\tau$ is constant across all time slots; 2) the
probability $P_{col}$ is constant and independent of the number of
collisions already suffered.

Given $\tau$, the probability of collision $P_{col}$, can be
defined as follows:
\begin{equation}
\label{eq:pcoll} P_{col} = 1 - (1-\tau)^{N-1}
\end{equation}
Finally, the term $q$ in (\ref{eq.tau}) is the probability of
having at least one packet waiting for transmission in the station
queue after an average slot duration.

Let us spend few words on the evaluation of $q$. Upon assuming
that the packet interarrival times are exponentially distributed
with mean\footnote{$\lambda$, which is measured in $pkt/s$,
represents the rate at which the packets arrive into the station
queue from the upper layers.} $1/\lambda$, the probability $q$ can
be well approximated by the following relation in a scenario where
the contending stations employ queues of small
sizes~\cite{hamilton}:
\begin{equation}%\small
\label{eq:q_prob}
\begin{array}{rlll} q &=& 1 - e^{- \lambda E[S_{ts}]}\approx \lambda\cdot E[S_{ts}]\left|_{\lambda\approx 0}\right.&\\
E[S_{ts}]&=&(1-P_t )\sigma+P_t(1-P_s)T_c+&\\
&&+P_tP_s(1-P_e)T_s+P_tP_sP_eT_e&
\end{array}
\end{equation}
where $E[S_{ts}]$, the \textit{expected time per slot}, is useful
to relate the states of the Markov chain to the actual time spent
in each state. As a note aside, notice that $e^{- \lambda
E[S_{ts}]}$ in~(\ref{eq:q_prob}) corresponds to the probability
that zero packets are received from the upper layers when the
packet interarrival times are exponentially distributed.

The other terms involved in $E[S_{ts}]$ are defined as
follows~\cite{Bianchi,daneshgaran}. $\sigma$ is the duration of an
empty time slot; $P_t$ is the probability that there is at least
one transmission in the considered time slot, with $N$ stations
contending for the channel, each transmitting with probability
$\tau$; $P_s$ is the conditional probability that a packet
transmission occurring on the channel is successful; $T_c$, $T_e$
and $T_s$ are, respectively, the average times a channel is sensed
busy due to a collision, the transmission time during which the
data frame is affected by channel errors, and the average time of
successful data frame transmission.
\section{Throughput Analysis}
\label{Section_Throughput_Analysis}
%%
%\figura{N10_m5_PL1024_nocapt_variPe_funz_lambda.eps}{Throughput
%for the 2-way mechanism as a function of the packet rate
%$\lambda$, for $N=10$, $m=5$, packet size $E[PL]=2312$ bytes, and
%different values of $P_e$ and $W_0$ as noticed in the respective
%legends. Continuous straight lines refer to the linear model of
%the throughput $S(\lambda)=N\cdot E[PL]\cdot \lambda$ derived
%in~(\ref{lin_model_throughput}), while the horizontal line
%corresponds to $S_m$ noticed in~(\ref{S_max_equat}). Dashed lines,
%labeled Thr, refer to the theoretical throughput in
%(\ref{eq.system2}), while dash-dotted lines labeled NS2 identify
%the simulation results obtained with
%ns-2.}{N10_m5_PL1024_nocapt_variPe_funz_lambda}
%%
%
The computation of the normalized system throughput relies on the
numerical solution of the nonlinear system obtained by jointly
solving~(\ref{eq.equ}) and~(\ref{eq.tau}). The solution of the
system, which corresponds to the values of $\tau$ and $P_{eq}$, is
used for the computation of the normalized system throughput,
i.e., the fraction of the time during which the channel is used to
successfully transmit payload bits:
\begin{equation}%\small
\label{eq.system2} S = P_t \cdot P_s\cdot (1-P_e)E[PL]/E[S_{ts}]
%}{E[S_{ts}]}
\end{equation}
whereby $E[PL]$ is the average packet payload length, and $P_t
\cdot P_s$ can be rewritten as \cite{Bianchi}:
\begin{equation}
\label{PtPs_equ} P_t \cdot P_s=N\tau (1-\tau)^{N-1}
\end{equation}
In order to gain insights on the behavior of the aggregate
throughput $S$, let us investigate the theoretical behavior of
(\ref{eq.system2}) as a function of the packet rate $\lambda$ for
two different values of the packet error probability $P_e$, and
minimum contention window $W_0=32$, in a scenario with $N=10$
contending stations transmitting at the bit rate $1$~Mbps.

The two rightmost subplots of
Fig.~\ref{N10_m5_PL1024_nocapt_variPe_funz_lambda} show the
theoretical behavior of the throughput in (\ref{eq.system2}), as
well as simulation results obtained with NS2 by employing the
typical MAC layer parameters for IEEE802.11b given in
Table~\ref{tab.design.times}~\cite{standard_DCF_MAC}. Other
parameters are noticed in the label of the figure. Let us spend a
few words about the simulation setup in ns-2. In order to account
for imperfect channel transmissions, the channel model is
implemented using the suggestions proposed in \cite{xiuchao},
where the outcomes of a binary, equiprobable random variable are
used to establish whether each packet is received erroneously. In
other words, the random variable is equal to $1$ with probability
$P_e$ (erroneous transmission), and $0$ with probability $1-P_e$.
We notice in passing that the theoretical model developed is
independent on the specific propagation channel. The only
parameter needed is $P_e$, which can be appropriately linked to
the specific wireless propagation channel upon specifying a
threshold of the signal-to-noise ratio allowing perfect reception
at the receiver \cite{proakis}.

We adopted the patch NOAH (\textit{NO Ad-Hoc}), available on the
authors' website\footnote{See
http://icapeople.epfl.ch/widmer/uwb/ns-2/noah/}, for emulating a
wireless network in infrastructure mode.

The employed traffic model is implemented by generating an
exponentially distributed random variable with expected value
$1/\lambda$, in accordance to the theoretical model developed in
Section \ref{Section_Problem_Formulation}.

Let us focus on the curves noticed in the two rightmost subplots
in Fig.~\ref{N10_m5_PL1024_nocapt_variPe_funz_lambda}. Basically,
there are two different operating regions of the throughput in
(\ref{eq.system2}). As $\lambda\rightarrow 0^+$, i.e., all the
contending stations approach unloaded traffic conditions, the
throughput can be approximated as a straight line passing through
the point $(S,\lambda)=(0,0)$:
\begin{equation}\label{lin_model_throughput}
S(\lambda)=N\cdot E[PL]\cdot\lambda
\end{equation}
This relation follows from the theoretical throughput noticed in
(\ref{eq.system2}) upon approximating the probabilities $P_t$ and
$P_s$ in the limit $\lambda\rightarrow 0^+$.

Indeed, as $\lambda\rightarrow 0^+$, the first relation in
(\ref{eq:q_prob}) yields $q\approx \lambda\cdot E[S_{ts}]$,
whereas the probability $\tau$ in (\ref{eq.tau}) can be well
approximated by $\tau\approx q/(1-P_{eq})$. Since
$P_{col}\rightarrow 0$ as $\lambda\rightarrow 0^+$ (because
$\tau\rightarrow 0$), it is $P_{eq}\rightarrow P_e$. Furthermore,
as $\lambda\rightarrow 0^+$, (\ref{PtPs_equ}) can be approximated
by the following relation:
\begin{equation}
\label{PtPs_equ_approx} P_t \cdot P_s\approx N\tau
=N\frac{q}{1-P_e}
\end{equation}
Upon substituting (\ref{PtPs_equ_approx}) in (\ref{eq.system2}),
the theoretical throughput can be well approximated as follows:
%
%%
%\begin{table}\caption{Typical network parameters}%\label{parameters}
%\begin{center}
%\begin{tabular}{c|c||c|c}\hline
%\hline MAC header & 24 bytes & Slot time $\sigma$ & 20 $\mu s$\\
%\hline PHY header & 16 bytes & SIFS & 10 $\mu s$\\
%\hline ACK & 14 bytes & DIFS & 50 $\mu s$\\
%\hline $\tau_p$ & $1\mu$s  &EIFS & 300 $\mu s$\\
%\hline $W_0$ & 32&  ACK timeout & 300 $\mu s$\\
%
%\hline m & 5 &CTS timeout & 300 $\mu s$\\
%
%\hline \hline
%\end{tabular}
% \label{tab.design.times}
%\end{center}
%\end{table}
%%\hline Payload size & 1024 bytes\\
%%\hline RTS & 20 bytes&EIFS & 300 $\mu s$\\
%%\hline CTS & 14 bytes\\
%%
\begin{eqnarray}\label{straight_line_model}
S(\lambda)\left|_{\lambda\approx 0^+}\right.&=&\frac{P_t \cdot
P_s\cdot (1-P_e)E[PL]}{E[S_{ts}]}\nonumber\\
&\approx& N E[PL]\frac{q}{E[S_{ts}]}\approx N\cdot
E[PL]\cdot\lambda
\end{eqnarray}
This linear model is depicted in the subplots of
Fig.~\ref{N10_m5_PL1024_nocapt_variPe_funz_lambda} overimposed to
both theoretical and simulated results. The key observation from
this result is that the aggregate throughput produced by $N$
stations approaching unloaded traffic conditions, is only
dependent on the number of stations, as well as on the packet
size. No other network parameters affect the aggregate throughput
in this operating region. Moreover, the results depicted in
Fig.~\ref{N10_m5_PL1024_nocapt_variPe_funz_lambda} denote that the
derived linear model is valid up to a critical value of $\lambda$
(identified by $\lambda_c$ throughout the paper), above which the
aggregate throughput no longer increases linearly with $\lambda$.
We note in passing that, given $\lambda_c$, there is no need to
simulate the network to obtain the aggregate throughput: it is
very well approximated by the theoretical relation derived in
(\ref{straight_line_model}) for any $\lambda\in [0,\lambda_c]$.

Once again, let us focus on the results shown in the rightmost
subplots of Fig.~\ref{N10_m5_PL1024_nocapt_variPe_funz_lambda}.
The aggregate throughput gets to a maximum at a proper value of
$\lambda$, above which the effect of the collisions among the
stations, as well as the propagation channel, let the throughput
reach a horizontal asymptote. The maximum value of $S(\lambda)$
over the whole range of values of $\lambda$ turns to be quite
useful for throughput optimization. Obtained by Bianchi
\cite{Bianchi} under the hypothesis of saturated network, and
considering only collisions among the stations, in our more
general model such a maximum can be estimated in two steps. First,
rewrite the throughput in (\ref{eq.system2}) as a function of
$\tau$ by using the relations that define $P_t$ and $P_s$ in terms
of the probability $\tau$. Then, equate to zero the derivative of
the throughput in (\ref{eq.system2}) with respect to $\tau$, and
obtain a solution identified by $\tau_m$. By doing so, the value
of $\tau$ for which the throughput gets maximized is easily
obtained:
%%
%\figura{lambda_PL.eps}{Behavior of $\lambda_c$ in
%(\ref{lambda_critici}) as a function of the packet size $E[PL]$,
%for two different values of $N$ as noticed in the respective
%legends. Other parameters are as follows: $W_0=32$, $m=5$, and
%$P_e=0$.}{lambda_PL}
%%
\begin{equation}%\small
\label{eq:tau_max}
    \tau_m = \frac{\sigma-\sqrt{\sigma\left[N\sigma - 2(N-1)(\sigma-T_c)\right] / N} }{(N-1)(\sigma-T_c)}
\end{equation}
Finally, evaluate the throughput in (\ref{eq.system2}) on the
solution $\tau_m$ noticed in (\ref{eq:tau_max}). Upon following
these steps, the maximum throughput takes on the following form
\begin{equation}%\small
\label{S_max_equat}
 S(\tau_m)= S_m =  \frac{E[PL]}{\left[T_s-\frac{T_c}{1-P_e}+\frac{T_e P_e}{1-P_e}\right] + \frac{(\sigma-T_c)(1-\tau_m)^N + T_c}{N\tau_m(1-\tau_m)^{N-1}(1-P_e)}}
\end{equation}
The critical value of $\lambda$ acts as a transition threshold
between two operating regions. In the first region, i.e., for
$\lambda\in [0,\lambda_c]$, the transmissions are affected by
relatively small equivalent error probabilities, $P_{eq}$. In this
operating region\footnote{In what follows, this operating region
will be identified by the acronym BLC, short for Below Link
Capacity region.}, collisions among stations occur rarely
($P_{col}$ is small) because of the reduced traffic load of the
contending stations, and the stations experience good channel
quality ($P_{e}$ is very small). For any $\lambda \leq \lambda_c$,
the network is not congested, and the $N$ contending stations are
able to transmit data below the link capacity limit denoted by
$S_m$. This is the reason for which the aggregate throughput grows
linearly with the traffic load $\lambda$.

On the other hand, the aggregate throughput tends to be
upperbounded by $S_m$ in (\ref{S_max_equat}) for $\lambda\ge
\lambda_c$. The reason is simple: the aggregate throughput in this
region\footnote{In what follows, this operating region will be
identified by the acronym LC, short for Link Capacity region.} is
affected either by the increasingly effects of the collisions
among the stations, or by worse channel conditions as exemplified
by $P_e\gg 0$. Further insights on this statement can be gained by
the results discussed below in connection with the curves shown in
Fig.~\ref{Thr_lambda_various_PL_N10_Pe_0}.

The critical value of $\lambda$ can be found as the abscissa where
the linear model of the throughput in (\ref{lin_model_throughput})
equates to $S_m$ in (\ref{S_max_equat}):
%
%%
%\figura{Thr_lambda_various_PL_N10_Pe_0.eps}{Behavior of the
%aggregate throughput as a function of $\lambda$, for five
%different values of the packet size as noticed in the legend. In
%each simulated scenario there are $N=10$ contending stations
%transmitting at 1 Mbps. Other parameters are as follows: $W_0=32$,
%$P_e=0$, and $m=5$. The values of $\lambda_c$ are noticed on the
%abscissa, whereas the horizontal lines, labeled by $S_m$,
%represent the maximum aggregate throughput (link capacity)
%evaluated through (\ref{S_max_equat}) in each considered scenario.
%All the thick curves refer to $E[PL]=128$ bytes, and various
%packet error probabilities as noticed in the
%figure.}{Thr_lambda_various_PL_N10_Pe_0}
%
\begin{equation}
\label{lambda_critici}
\begin{array}{rl}
  \lambda_c  = & \left[N\left(T_s-\frac{T_c}{1-P_e}+\frac{T_e P_e}{1-P_e}\right) + \frac{(\sigma-T_c)(1-\tau_m)^N + T_c}{\tau_m(1-\tau_m)^{N-1}(1-P_e)}\right]^{-1}
\end{array}
\end{equation}
The procedure for obtaining $\lambda_c$, is clearly highlighted in
all the four subplots of
Fig.~\ref{N10_m5_PL1024_nocapt_variPe_funz_lambda}.

Let us investigate the behavior of the critical value $\lambda_c$
in (\ref{lambda_critici}) against some key network parameters.

Fig.~\ref{lambda_PL} shows the behavior of $\lambda_c$ as a
function of the packet size $E[PL]$, for two different values of
$N$ as noticed in the respective legends. Some observations are in
order. $\lambda_c$ decreases for longer packet sizes $E[PL]$, as
well as for increasing number of contending stations, $N$. The
reason relies on the fact that for increasing packet sizes, each
station tends to occupy the channel longer. Such a behavior is
clearly emphasized in Fig.~\ref{Thr_lambda_various_PL_N10_Pe_0},
where the aggregate throughput for a sample scenario comprising 10
contending stations transmitting with the packet sizes noticed in
the legend, is considered.

As long as the packet size increases, the aggregate throughput
shows an increasing slope in the linear region characterized by
traffic loads $\lambda\in [0,\lambda_c]$, as suggested by the
theoretical model in (\ref{straight_line_model}). Moreover, the
value of $\lambda_c$ tends to decrease because each station tends
to occupy the channel longer. The three lower thick curves,
labeled by $PL=128$ bytes, are associated to three different
values of packet error probabilities, $P_e$. Note that the value
of $\lambda_c$ decreases for increasing values of $P_e$, thus
reducing the region characterized by small equivalent error
probabilities. As long as $P_e\rightarrow 1$, both $\lambda_c$ and
$S_m$ tend to zero, making the aggregate throughput vanishingly
small.

Similar considerations may be derived from the behavior of
$\lambda_c$ versus $N$ noticed in Fig.~\ref{lambda_N}. Roughly
speaking, for fixed values of the packet size, $\lambda_c$ tends
to reduce by a half as far as the number of contending stations
$N$ doubles. The reason relies on the fact that the probability of
collision increases as long as more stations try to contend for
the channel.
%%
%\figura{lambda_N.eps}{Behavior of $\lambda_c$ in
%(\ref{lambda_critici}) as a function of the number of contending
%stations, $N$, for two different values of $E[PL]$ as noticed in
%the respective legends. Other parameters are as follows: $W_0=32$,
%$m=5$, and $P_e=0$.}{lambda_N}
%%
%
%
%
\section{Throughput Optimization}
\label{Section_throughput_Optimization}
The considerations deduced in Section
\ref{Section_Throughput_Analysis} are at the very basis of an
optimization strategy for maximizing the aggregate throughput of
the network depending on the traffic load $\lambda$. To this end,
we define two optimization strategies: \emph{Contention Window
Optimization} and \emph{MAC Payload Size Optimization}. The first
strategy is applied when the contending stations operate in the LC
region, approaching the link capacity $S_m$, whereas the second
one is used for optimizing the aggregate throughput when the
stations operate within the BLC region.

The next two subsections address separately the two optimization
strategies, while Section \ref{Section_simulation_results}
presents the optimization algorithm jointly implementing the two
strategies.
\subsection{Link Capacity region: Contention Window Optimization}
\label{Section_throughput_OptimizationHEEPR_region}
The first optimization strategy proved to be effective for
improving the aggregate throughput in the LC region, i.e., for
$\lambda>\lambda_c$. The key idea here is to force the contending
stations to transmit with a probability $\tau$ equal to the one
that maximizes the aggregate throughput $S(\lambda)$. In this
respect, the probability $\tau_m$ in (\ref{eq:tau_max}) plays a
key role.

Upon considering saturated conditions, i.e., imposing
$q\rightarrow 1$ in (\ref{eq.tau}), the probability $\tau$ can be
rewritten as:
\begin{equation}\label{eq.tau_2}
\begin{array}{lll}
  \tau & = & \frac{2(1-2P_{eq})}{(W_0+1)(1-2P_{eq}) + W_0P_{eq}(1-(2P_{eq})^m)}
\end{array}
\end{equation}
%
%For the sake of simplifying the notation, let us rewrite $P_{eq}$
By substituting $\tau=\tau_m$ in~(\ref{eq:pcoll}), $P_{eq}$ can be
rewritten as:
\begin{equation}%\small
\label{eq.peq_1}
\begin{array}{lll}
P_{eq}&=&1-(1-\tau_m)^{N-1}+P_e-P_e [1-(1-\tau_m)^{N-1}]\\
 &=& 1+(P_e-1)(1-\tau_m)^{N-1}=1-X(P_e,\tau_m)
\end{array}
\end{equation}
whereby $X(P_e,\tau_m)=(1-P_e)(1-\tau_m)^{N-1}$.

Finally, by equating~(\ref{eq.tau_2}) to $\tau_m$
in~(\ref{eq:tau_max}), and solving for $W_0$, we obtain the
optimal minimum contention window size in terms of the key network
parameters:
\begin{equation}%\small
\label{W_optimal}
\begin{array}{lll}
W_{OP}&=&\frac{1-2\tau_m^{-1}+X(P_e,\tau_m)\left(4\tau_m^{-1}-2\right)}{2X(P_e,\tau_m)-1+\left(1-X(P_e,\tau_m)\right)\left[1-2^m\left(1-X(P_e,\tau_m)\right)^m\right]}
\end{array}
\end{equation}
This relation yields the value of the minimum contention window
$W_0$ that maximizes the aggregate throughput when the number of
contending stations $N$, the packet error rate over the channel
$P_e$, and the number of backoff stages $m$, are given. As a note
aside, notice that, using $W_{OP}$, the maximum throughput equates
to the link capacity $S_m$ in~(\ref{S_max_equat}).

Moreover, we notice that for $m=0$, i.e., no exponential backoff
is employed, $W_{OP}$ in (\ref{W_optimal}) can be simplified as
follows:
%
%%
%\figura{N10_m5_PL1024_nocapt_variPe.eps}{Saturation throughput
%(continuous curves) as a function of the minimum contention window
%size $W_0$, for three different values of $P_e$ as noticed in the
%legend. The other parameters are as follows: $N=10$, $m=5$, and
%$E[PL]=1024$ bytes. The dashed curves represent the theoretical
%throughput in (\ref{eq.system2}) for the chosen set of parameters.
%Notice that such curves are superimposed to the results obtained
%with NS-2.}{N10_m5_PL1024_nocapt_variPe}
%%
\[
W_{OP}=\frac{2}{\tau_m}-1
\]
Considering $\tau_m$ as a function of $N$, and neglecting the
multiplicative constant terms, it is simple to notice that
$\tau_m$ in (\ref{eq:tau_max}) goes roughly as $1/N$ when $N\gg
1$. Therefore, the optimal contention window $W_{OP}$ grows
linearly with the number of contending stations $N$ (when $N\gg
1$) in order to mitigate the effects of the collisions due to an
increasing number of contending stations in the network.

In the following, we present simulation results accomplished in
NS-2 for validating the theoretical models, as well as the results
presented in this section. The adopted MAC layer parameters for
IEEE802.11b are summarized in
Table~\ref{tab.design.times}~\cite{standard_DCF_MAC}.

The main simulation results are presented in Figs.
\ref{N10_m5_PL1024_nocapt_variPe_funz_lambda} and
\ref{N10_m5_PL1024_nocapt_variPe} in connection to the set of
parameters noticed in the respective labels.

Some observations are in order. Let us focus on the results shown
in Fig.~\ref{N10_m5_PL1024_nocapt_variPe_funz_lambda}. A quick
comparison among the leftmost and the rightmost subplots of
Fig.~\ref{N10_m5_PL1024_nocapt_variPe_funz_lambda} reveals that
the choice $W_0=W_{OP}$ in (\ref{W_optimal}) guarantees improved
performance for any $\lambda>\lambda_c$, thus making the aggregate
throughput equal $S_m$. Throughput penalties due to the use of an
suboptimal $W_0$, are in the order of $100$ kbps.

Notice also that the value of $\lambda_c$ is independent from the
minimum contention window chosen.

The key observation from the subplots of
Fig.~\ref{N10_m5_PL1024_nocapt_variPe_funz_lambda} concerns the
fact that the aggregate throughput of the optimized network can be
modeled as follows:
 \[
 S(\lambda)\left|_{W_{OP}}\right.=\left\{
\begin{array}{ll}
N\cdot E[PL]\cdot \lambda, & \lambda\le \lambda_c\\
S_m, & \lambda> \lambda_c
\end{array}\right.
 \]
 As emphasized by the curves in Fig.~\ref{N10_m5_PL1024_nocapt_variPe_funz_lambda},
 this model shows a very good agreement with both theoretical and simulation
 results. Once again, notice that the aggregate throughput may be predicted quite
 accurately without resorting to simulation.

Let us focus on the results shown in
Fig.~\ref{N10_m5_PL1024_nocapt_variPe}, where the saturation
throughput represented by continuous curves, is derived as a
function of the minimum contention window size $W_0$. The curves
are parameterized with respect to the three different values of
$P_e$ noticed in the legend. The simulated scenario considers
$N=10$ contending stations transmitting packets of size
$E[PL]=1024$ bytes. Shown in the same figure is the theoretical
throughput in (\ref{eq.system2}) represented by dashed curves.

The saturation throughput in each simulated scenario reaches a
maximum corresponding to the abscissas $W_0=W_{OP}$ predicted by
(\ref{W_optimal}) and noticed beside each maximum. We notice that
$W_{OP}$ tends to be quite insensitive from the packet error rate
$P_e$, whose main effect corresponds to a reduction of the maximum
achievable throughput.
\subsection{Below Link Capacity region: Payload Size Optimization}
\label{Section_throughput_OptimizationLEEPR_region}
The analysis of the aggregate throughput in Section
\ref{Section_Throughput_Analysis} revealed the basic fact that for
traffic loads $\lambda$ less than $\lambda_c$, the network is not
congested, and each station achieves a throughput roughly equal to
$E[PL]\cdot \lambda$ (the aggregate throughput is thus
$S(\lambda)=N\cdot E[PL]\cdot \lambda$). We note in passing that
the throughput does not depend on the minimum contention window
$W_0$ in the LC region. Therefore, given $N$ contending stations,
the throughput can only be improved by increasing the payload size
$E[PL]$ when the traffic load satisfies the relation $\lambda\le
\lambda_c$.

Before proceeding any further, let us discuss two important issues
in connection with the choice of the packet size $E[PL]$.

As long as the erroneous bits are independently and identically
distributed over the received packet\footnote{We notice that
wireless transceivers make use of interleaving in order to break
the correlation due to the frequency selectivity of the
transmission channel \cite{proakis}.}, the packet error
probability $P_e$ can be evaluated as
\begin{equation}\label{fer_1}
P_e=1-\left[1-P_e(PLCP)\right]\cdot \left[1-P_e(DATA)\right]
\end{equation}
where,
\begin{equation}\label{fer_2}
P_e(PLCP)=1-\left[1-P_b(BPSK)\right]^{PHYh},
\end{equation}
and
\begin{equation}\label{fer_3}
P_e(DATA)=1-(1-P_b)^{MACh+E[PL]}
\end{equation}
In the previous relations, $MACh$ and $PHYh$ are, respectively,
the sizes (in bits) of the MAC and PLCP\footnote{PLCP is short for
Physical Layer Convergence Protocol.} headers, and $E[PL]$ is the
size of the data payload in bits. Equ. (\ref{fer_3}) accounts for
the fact that a packet containing the useful data, is considered
erroneous when at least one bit is erroneously received. Relation
(\ref{fer_1}) is obtained by noting that a packet is received
erroneously when the errors occur either in the PLCP part of the
packet, or in the information data.

We notice that the relations (\ref{fer_1}) through (\ref{fer_3})
are valid either with the use of convolutional coding, whereby the
bit error rate performance does not depend on the code block size
\cite{proakis}, or for protocols that do not employ channel
encoding at the physical layer.

From (\ref{fer_3}) it is quite evident that the higher the payload
size, the higher the packet error rate; and viceversa. As a note
aside, we notice that this behavior does not hold when
concatenated convolutional channel codes
\cite{laddomada_5}-\cite{laddomada_4}, as well as low-density
parity-check codes, are employed as channel codes. Indeed, for
these codes the packet error rate depends on the size of the
encoded block of data, with longer packets having smaller
probability of error compared to shorter packets.

The second issue to be considered during the choice of $E[PL]$ is
related to the critical value $\lambda_c$, which depends on the
packet error probability ($P_e$ defined in (\ref{fer_1})), and
consequently on the payload size. Given a traffic load $\lambda$,
any change in $E[PL]$ could affect the network operating region,
which might move from the BLC region to the LC region.

Let us discuss a simple scenario in order to reveal this
issue\footnote{Where not otherwise specified, we employ the
network parameters summarized in Table \ref{tab.design.times}.}.
Consider a network (identified in the following as scenario A)
where $N=10$ contending stations with traffic load
$\lambda=8$~pkt/s transmit packets of size $E[PL]=1024$~bytes.
Assume that the bit error probability due to the channel
conditions is $P_b=10^{-5}$, which corresponds to a packet error
rate $P_e=8.248\cdot 10^{-2}$ (from (\ref{fer_1})). Upon using the
network parameters summarized in Table \ref{tab.design.times}, the
critical load in (\ref{lambda_critici}) corresponds to
$\lambda_c=9.61$~pkt/s. Since $\lambda<\lambda_c$, the network is
in the BLC operating region. If the stations increase the payload
size up to $E[PL]=2048$~bytes, the packet error rate increases to
the value $P_e=1.546\cdot 10^{-1}$, with the side effect of
decreasing $\lambda_c$ to $4.71$~pkt/s. Therefore, for the given
traffic load $\lambda=8$~pkt/s, the network starts operating into
the LC region.

Based on the considerations deduced in the scenario A, the
proposed optimization technique aims to optimize the payload size
in two consecutive steps. In the first step, we find the size of
the payload in such a way that the critical threshold $\lambda_c$
equals the actual traffic load $\lambda$, in order for the network
not to operate beyond the BLC region.

The second step verifies whether the packet error rate associated
to this payload size, is below a predefined PER-target (identified
by PER$_t$), which defines the maximum error level imposed by the
application layer. First of all, given an estimated bit error
probability at the physical layer, solve (\ref{fer_3}) for
$E[PL]$. Then, solve (\ref{fer_1}) for $1-P_e(DATA)$, and
substitute the relation in place of $1-P_e(DATA)$. Considering
$P_e=$PER$_t$, the following maximum payload size follows:
\begin{equation}\label{eq:FER_PL}
  E[PL]^{max}\mid_{PER_t}=\left\lceil \frac{\ln(\frac{1-PER_t}{1-P_e(PLCP)})}{\ln(1-P_b)} - MACh \right\rceil
\end{equation}
whereby $\lceil\cdot\rceil$ is the ceil of the enclosed number.

Let us consider again the scenario A discussed previously with the
application of this algorithm. Consider $10$ stations transmitting
packets of size $E[PL]=1024$ bytes at the traffic load $\lambda=5$
pkt/s. Assume that the bit error rate imposed by the specific
channel conditions is $P_b=10^{-5}$, yielding $P_e=8.248\cdot
10^{-2}$ and $\lambda_c=9.6$ pkt/s. Moreover, assume a target PER
equal to\footnote{This is the maximum PER specified in the IEEE
802.11b standard \cite{standard_DCF_MAC}, guaranteed at the
receiver when the received power reaches the receiver
sensitivity.} PER$_t=8 \cdot 10^{-2}$.

Since $\lambda<\lambda_c$, after the first step, the algorithm
selects the optimal size $E[PL]^*_1 = 1938$ bytes, which moves the
working point near the link capacity and leads to the actual
packet error rate $P_e = 1.47 \cdot 10^{-1}$. Since the constraint
$P_e \leq$PER$_t$ is not satisfied, the proposed method estimates
the size $E[PL]$ in order to attain the PER$_t$ constraint. The
packet size becomes eventually $E[PL]^*_2=991$ bytes (from
(\ref{eq:FER_PL})), yielding a packet error rate $P_e \simeq 8
\cdot 10^{-2}$ and a $\lambda_c^*=9.92$ pkt/s, thus leaving the
working point in the BLC region.

Finally, the optimal payload size is: $$E[PL]=\min\{E[PL]_1^*,
E[PL]_2^*,PL_{max}\}=E[PL]_2^*$$
where $PL_{max}=2312$ bytes is the maximum packet size imposed by
the standard \cite{standard_DCF_MAC}.

Simulation results of a sample network employing the algorithm
described above, are presented in the next section, along with a
sample code fragment summarizing the key steps of the optimization
technique.
\section{Simulation Results}
\label{Section_simulation_results}
%%
%\begin{table}
%  \centering\label{Optimization Algorithm}
%  \caption{Optimization algorithm accomplished by every contending station.}
%  \begin{tabular}{rl}
%    %\hline
%    % after \\: \hline or \cline{col1-col2} \cline{col3-col4} ...
%    %\multicolumn{2}{l}{Optimization Algorithm}\\
%    \hline
%    1. & \verb"do" \\
%    2. & \qquad Estimation of PER and N  \\
%    3. & \qquad Evaluate $\lambda_c$ from (\ref{lambda_critici})\\
%    4. & \verb"while(IDLE)" \\
%    5. & // Request to Send from Upper Layers:\\
%    6. & \verb"if ("$\lambda > \lambda_c$\verb")" \\
%    7. & \qquad // Link Capacity Operating Region: \\
%    8. & \qquad The station evaluates $W_{opt}$ from (\ref{W_optimal}) \\
%    9. & \verb"else" \\
%    10. & \qquad // Below Link Capacity Operating Region:\\
%    11. &\qquad compute $E[PL]^*_{1}$ given $\lambda_c$ \\
%    12. & \qquad compute $E[PL]^*_{2}$ given $P_e$ \\
%    13. &  \qquad select $E[PL]_{opt} = \min(E[PL]^*_{1},E[PL]^*_{2}, PL_{max})$ \\
%    14. & \verb"end" \\
%    15. & send packet \\
%    \hline
%  \end{tabular}
%\end{table}
%%
%
In this section, we present simulation results for a network of
$N=10$ contending stations, employing the optimization strategies
described in Section \ref{Section_throughput_Optimization}.

The basic steps of the proposed optimization algorithm are
summarized in Table \ref{Optimization Algorithm}. Let us spend a
few words about the implementation of the proposed algorithm in a
WLAN setting employing the infrastructure mode, where an Access
Point (AP) monitors the transmissions of $N$ contending stations.

The evaluation of the critical value $\lambda_c$ in
(\ref{lambda_critici}) requires the estimation of three key
parameters of the network: the number of contending stations $N$,
the packet error rate $P_e$, and the value of the optimal
$\tau_m$, which depends on $N$. Each station can estimate the
number of contending stations by resorting to one of the
algorithms proposed in \cite{toledo}-\cite{Serpedin}. On the other
hand, the packet error probability $P_e$, may be evaluated through
(\ref{fer_1}), once the bit error probability $P_b$ at the
physical layer is estimated. We recall that $P_b$ can be estimated
by employing proper training sequences at the physical layer of
the wireless receiver \cite{proakis}.

Finally, the value of the optimal $W_{OP}$ can be evaluated by
each station with the parameters $N$ and $P_e$. A similar
reasoning applies for the estimation of the optimal packet size.
%%
%\figura{final_simulation.eps}{Aggregate throughput of a network
%with a maximum number of stations equal to 10, transmitting at the
%fixed bit rate 1Mbps. The other network parameters are described
%in the paper.}{final_simulation}
%%

We have realized a C++ simulator implementing all the basic
directives of the IEEE 802.11b protocol with the 2-way handshaking
mechanism, namely exponential backoff, waiting times,
post-backoff, and so on. In our simulator, the optimization
algorithm is dynamically executed for any specified scenario. For
the sake of analyzing the effects of the optimization, the
instantaneous network throughput is evaluated over the whole
simulation. The aggregate throughputs obtained in the investigated
scenarios are shown in Fig.~\ref{final_simulation}.

In the first scenario, we considered a congested network in which
10 stations transmit packet of fixed size $1028$ bytes at the
packet load $\lambda=1000$ pkt/s and bit rate 1 Mbps. Other
parameters are $W_0=32$, and $m=5$, whereas the channel conditions
are assumed to be ideal. As shown in Fig.~\ref{final_simulation}
(curve labeled LC region with asterisk-marked points), the
aggregate throughput is about $7.6\times 10^{5}$ bps. After $40$s,
$5$ out of the $10$ stations turn their traffic off. Between 40s
and 80s, the aggregate throughput increases to about $8.2\times
10^{5}$ bps because of the reduced effect of the collisions among
the contending stations. After $80$s, the $5$ stations turn on
again and the aggregate throughput decreases to $7.6\times 10^{5}$
bps.

Consider again the same scenario over a time interval of $120$s,
whereby the contending stations adopt the optimal contention
window. Upon using (\ref{W_optimal}), the optimal contention
window is $W_{OP}^{0-40}=275$ when 10 stations transmit over the
network, and $W_{OP}^{40-80}=130$ during the interval in which
only 5 stations contend for the channel.

The aggregate throughput in the optimized scenario is about
$8.6\times 10^{5}$ bps, as noticed by the star-marked curve in the
same figure. We notice that the optimized throughput is quite
constant over the whole simulation independently on the number of
contending stations in the network, approaching the theoretical
maximum throughput $S_m=8.6\times 10^{5}$ bps obtained from
(\ref{S_max_equat}). Moreover, notice that the two curves $S_m$
related to both $N=5$ and $10$ are almost superimposed.

Consider another scenario differing from the previous one in that
the contending stations operate in the BLC region. As above, $10$
stations contend for the channel in the time intervals $[0,40]$s
and $[80,120]$s, while during the time interval $[40,80]$s $5$ out
of the 10 stations turn off. The traffic load is $\lambda=8$
pkt/s, and the packet inter-arrival times are exponentially
distributed.

Let us focus on the aggregate throughput depicted in
Fig.~\ref{final_simulation}. The curves related to the scenario at
hand are identified by the labels $\lambda=8$ pkt/s. As a result
of the application of the proposed algorithm to the choice of the
packet size, we obtain $E[PL]_{opt}^{0-40}=1383$ bytes for the
case when $10$ stations are transmitting, and
$E[PL]_{opt}^{40-80}=2312$ bytes in the other scenario with $5$
active stations.

Simulation results show that the proposed algorithm guarantees
improved throughput performance on the order of $160$ kbps when
$10$ stations are active, and about $400$ kbps when only $5$
stations contend for the channel. We notice in passing that,
despite the optimization, the aggregate throughput could not reach
the maximum $S_m$ because of the low traffic load. Indeed, the
capacity link $S_m$ could be achieved by using a packet size
longer than either the one imposed by the standard
\cite{standard_DCF_MAC}, and the value obtained from
(\ref{eq:FER_PL}).

For the sake to verify that the optimization of the minimum
contention window does not affect the aggregate throughput in the
BLC region, Fig.~\ref{final_simulation} also shows the aggregate
throughput obtained by simulation using a minimum contention
window equal to $W_{OP}$. Notice that the related curve is
superimposed to the one related to the scenario in which the
contending stations employ the minimum contention window $W_0=32$
suggested by the standard \cite{standard_DCF_MAC}.
\section{Conclusions}
\label{Section_Conclusions}
This paper proposed an optimization framework for maximizing the
throughput of the Distributed Coordination Function (DCF) basic
access mechanism at the data link layer of IEEE 802.11 protocols.
Based on the theoretical derivations, as well as on simulation
results, a simple model of the optimized DCF throughput has been
derived. Such a model turns to be quite useful for predicting the
aggregate throughput of the DCF in a variety of network
conditions.

For throughput modeling, we considered general operating
conditions accounting for both non-saturated and saturated traffic
in the presence of transmission channel errors identified by the
packet error rate $P_e$. Simulation results closely matched the
theoretical derivations, confirming the effectiveness of both the
proposed DCF model and the cross-layer optimization algorithm.
%
%\clearpage
%

%
\clearpage
\figuramedia{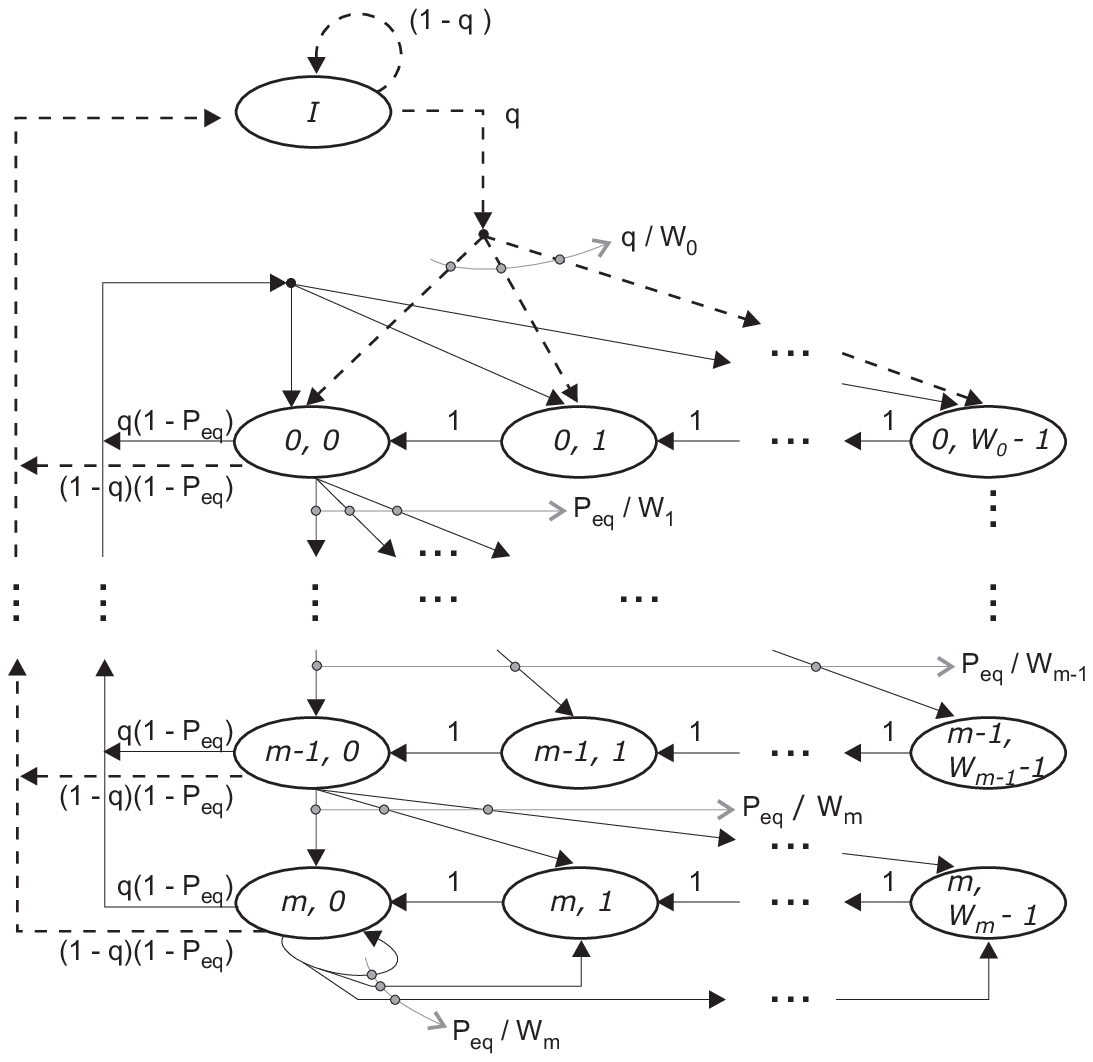}{Markov chain for the
contention model accounting for unsaturated traffic conditions.
The model considers $m$ backoff stages. $P_{eq}$ is the
probability of failed transmission, accounting for both collisions
and channel errors affecting the transmitted packets.}{fig.chain}
\clearpage
\figura{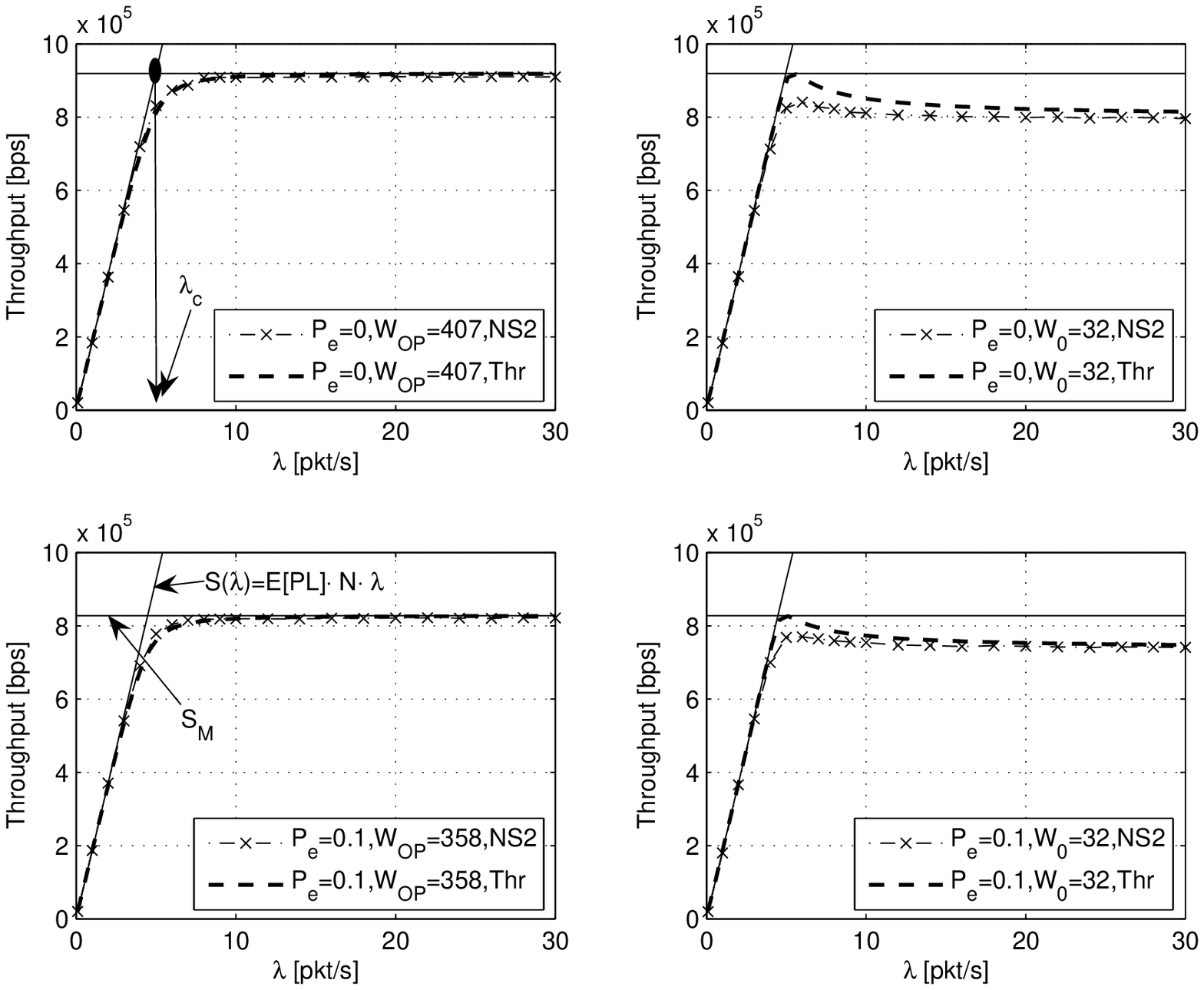}{Throughput
for the 2-way mechanism as a function of the packet rate
$\lambda$, for $N=10$, $m=5$, packet size $E[PL]=2312$ bytes, and
different values of $P_e$ and $W_0$ as noticed in the respective
legends. Continuous straight lines refer to the linear model of
the throughput $S(\lambda)=N\cdot E[PL]\cdot \lambda$ derived
in~(\ref{lin_model_throughput}), while the horizontal line
corresponds to $S_m$ noticed in~(\ref{S_max_equat}). Dashed lines,
labeled Thr, refer to the theoretical throughput in
(\ref{eq.system2}), while dash-dotted lines labeled NS2 identify
the simulation results obtained with
ns-2.}{N10_m5_PL1024_nocapt_variPe_funz_lambda}
\clearpage
\figura{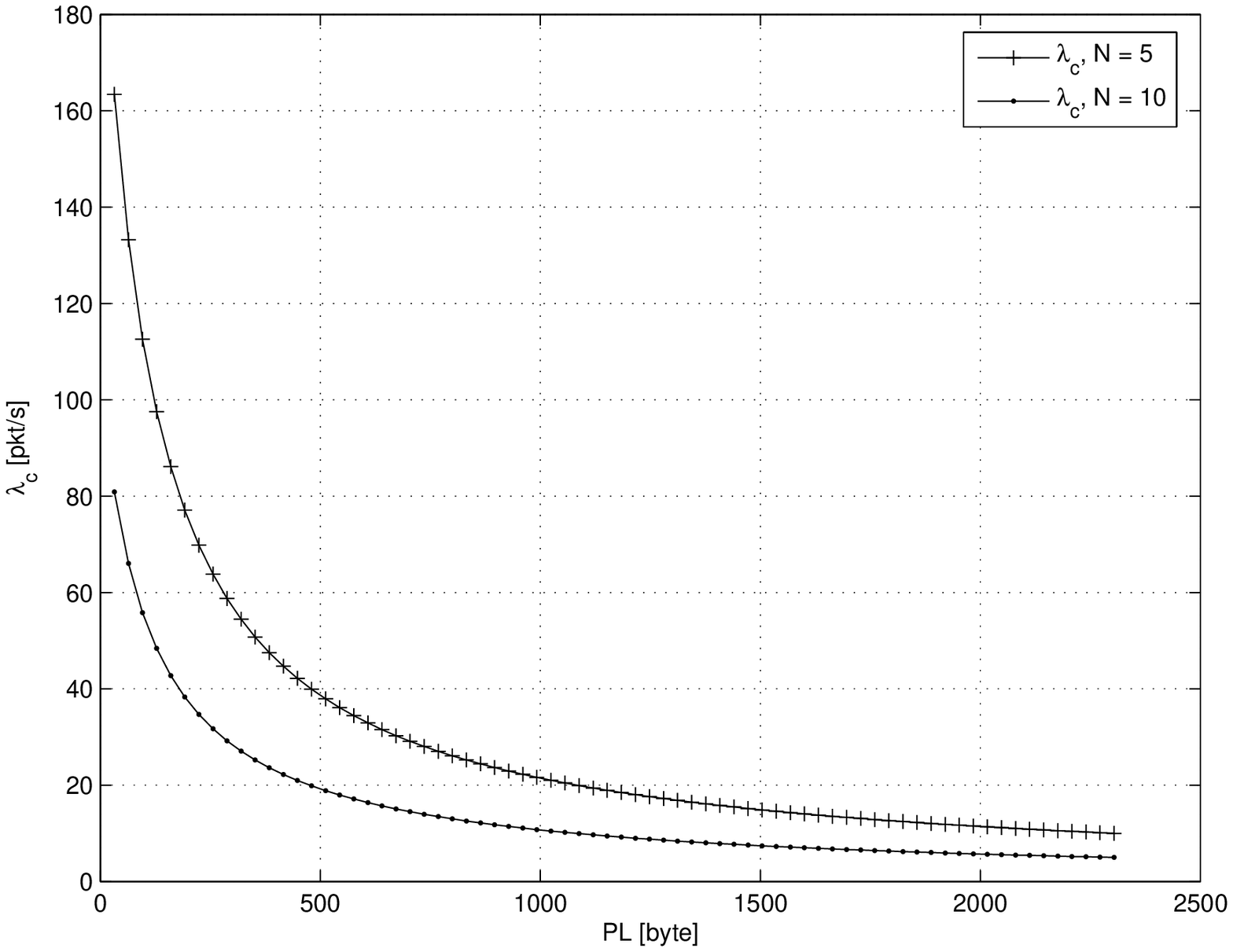}{Behavior of $\lambda_c$ in
(\ref{lambda_critici}) as a function of the packet size $E[PL]$,
for two different values of $N$ as noticed in the respective
legends. Other parameters are as follows: $W_0=32$, $m=5$, and
$P_e=0$.}{lambda_PL}
\clearpage
\figura{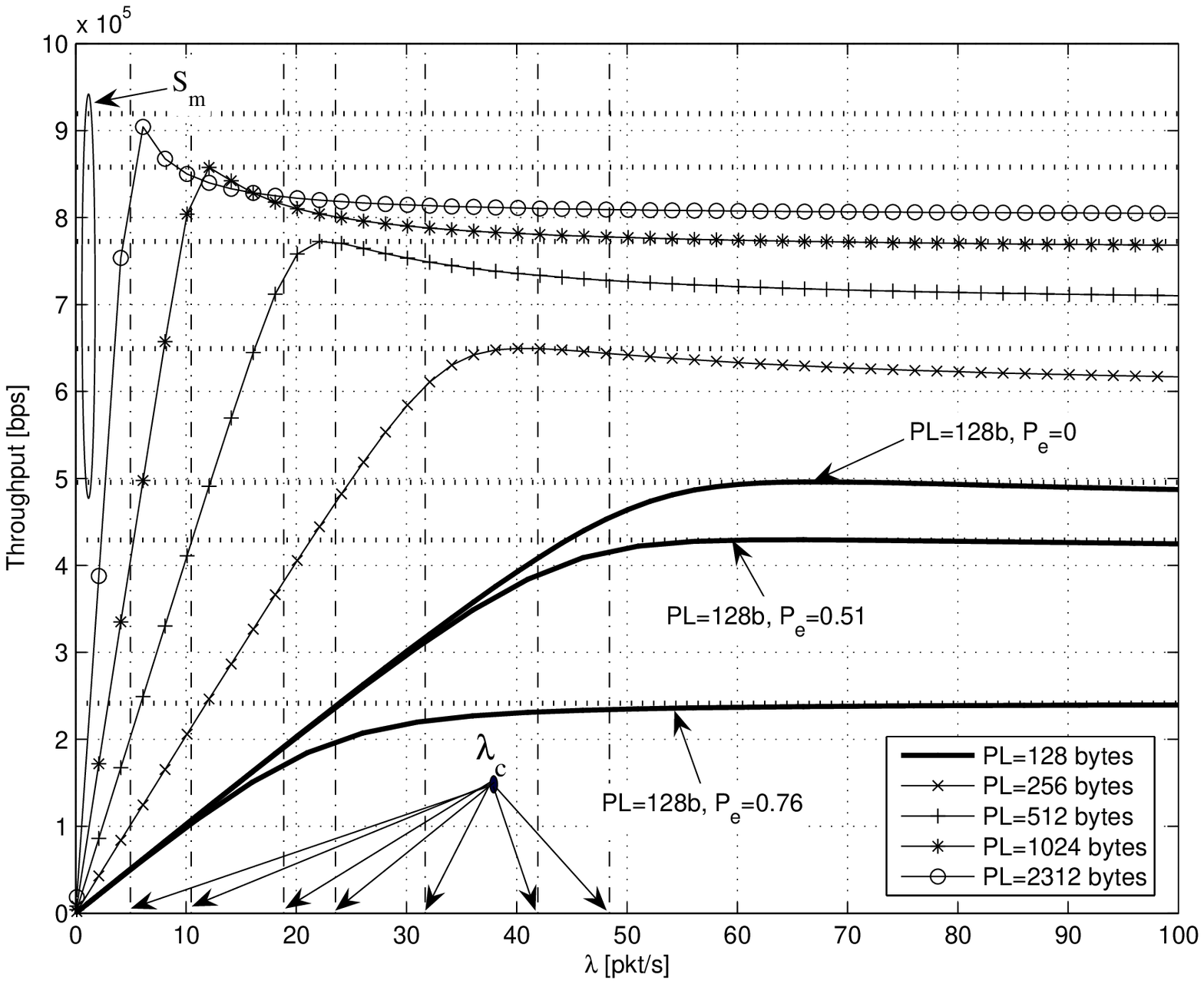}{Behavior of the
aggregate throughput as a function of $\lambda$, for five
different values of the packet size as noticed in the legend. In
each simulated scenario there are $N=10$ contending stations
transmitting at 1 Mbps. Other parameters are as follows: $W_0=32$,
$P_e=0$, and $m=5$. The values of $\lambda_c$ are noticed on the
abscissa, whereas the horizontal lines, labeled by $S_m$,
represent the maximum aggregate throughput (link capacity)
evaluated through (\ref{S_max_equat}) in each considered scenario.
All the thick curves refer to $E[PL]=128$ bytes, and various
packet error probabilities as noticed in the
figure.}{Thr_lambda_various_PL_N10_Pe_0}
\clearpage
\figura{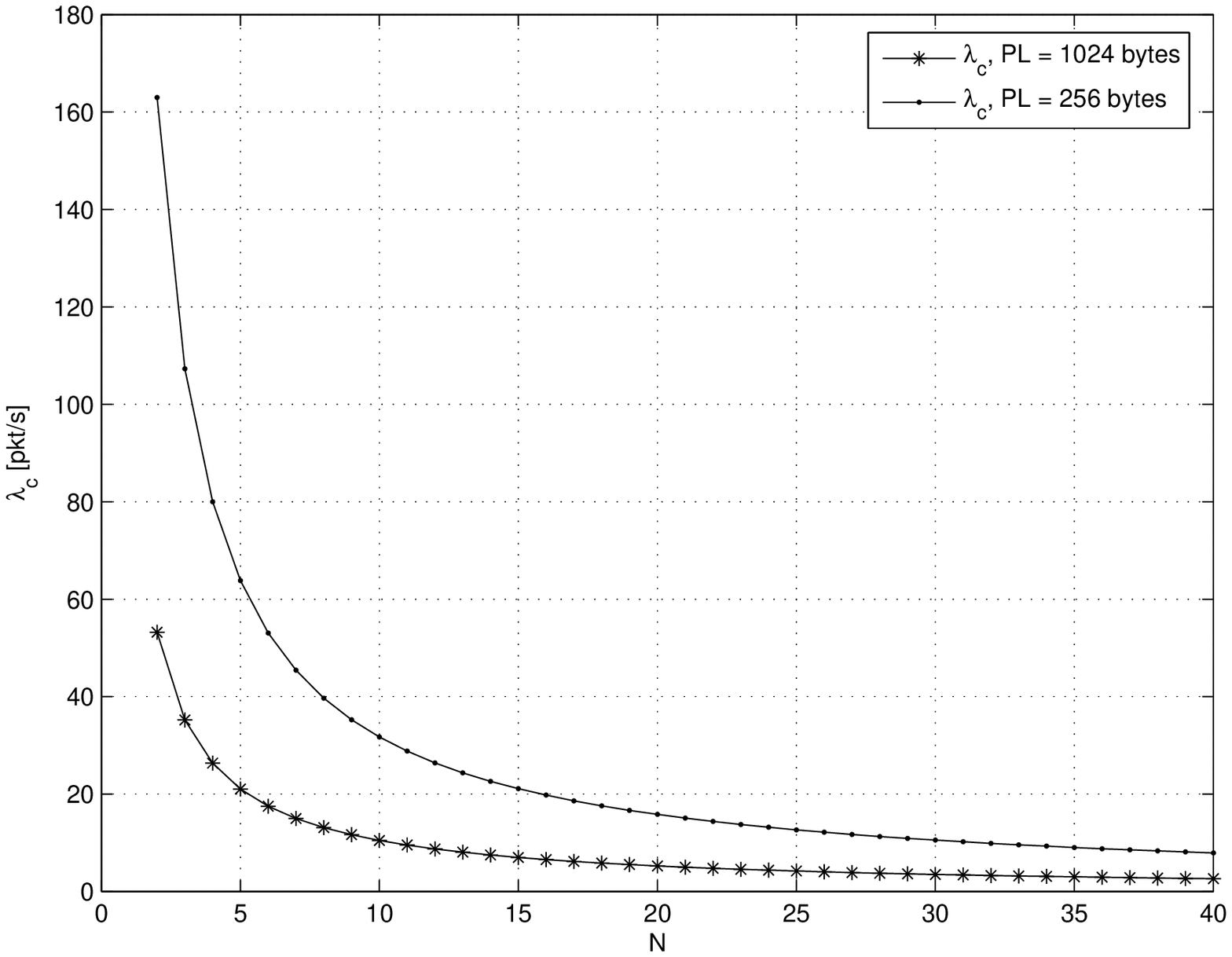}{Behavior of $\lambda_c$ in
(\ref{lambda_critici}) as a function of the number of contending
stations, $N$, for two different values of $E[PL]$ as noticed in
the respective legends. Other parameters are as follows: $W_0=32$,
$m=5$, and $P_e=0$.}{lambda_N}
\clearpage
\figura{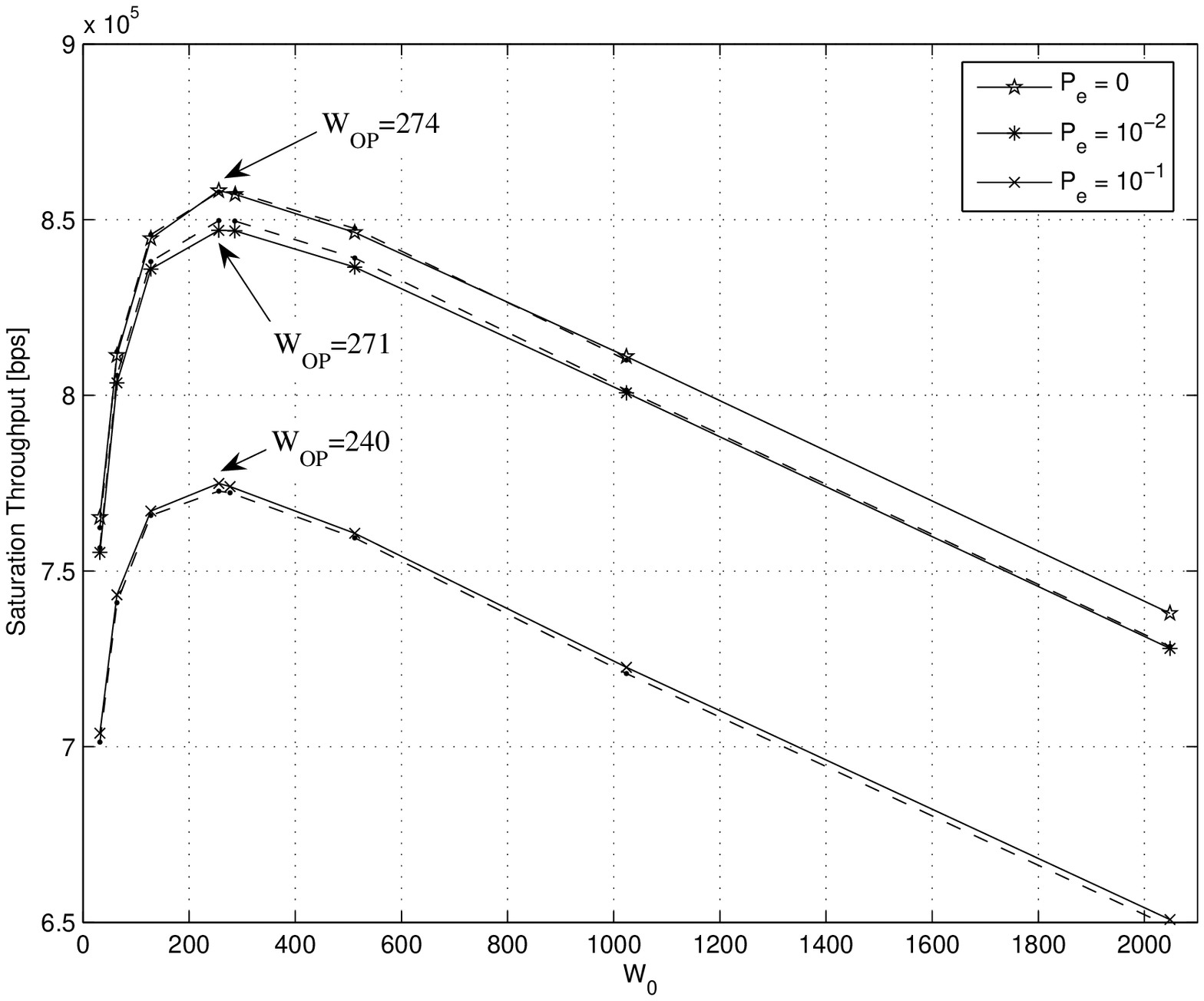}{Saturation throughput
(continuous curves) as a function of the minimum contention window
size $W_0$, for three different values of $P_e$ as noticed in the
legend. The other parameters are as follows: $N=10$, $m=5$, and
$E[PL]=1024$ bytes. The dashed curves represent the theoretical
throughput in (\ref{eq.system2}) for the chosen set of parameters.
Notice that such curves are superimposed to the results obtained
with NS-2.}{N10_m5_PL1024_nocapt_variPe}
\clearpage
\figura{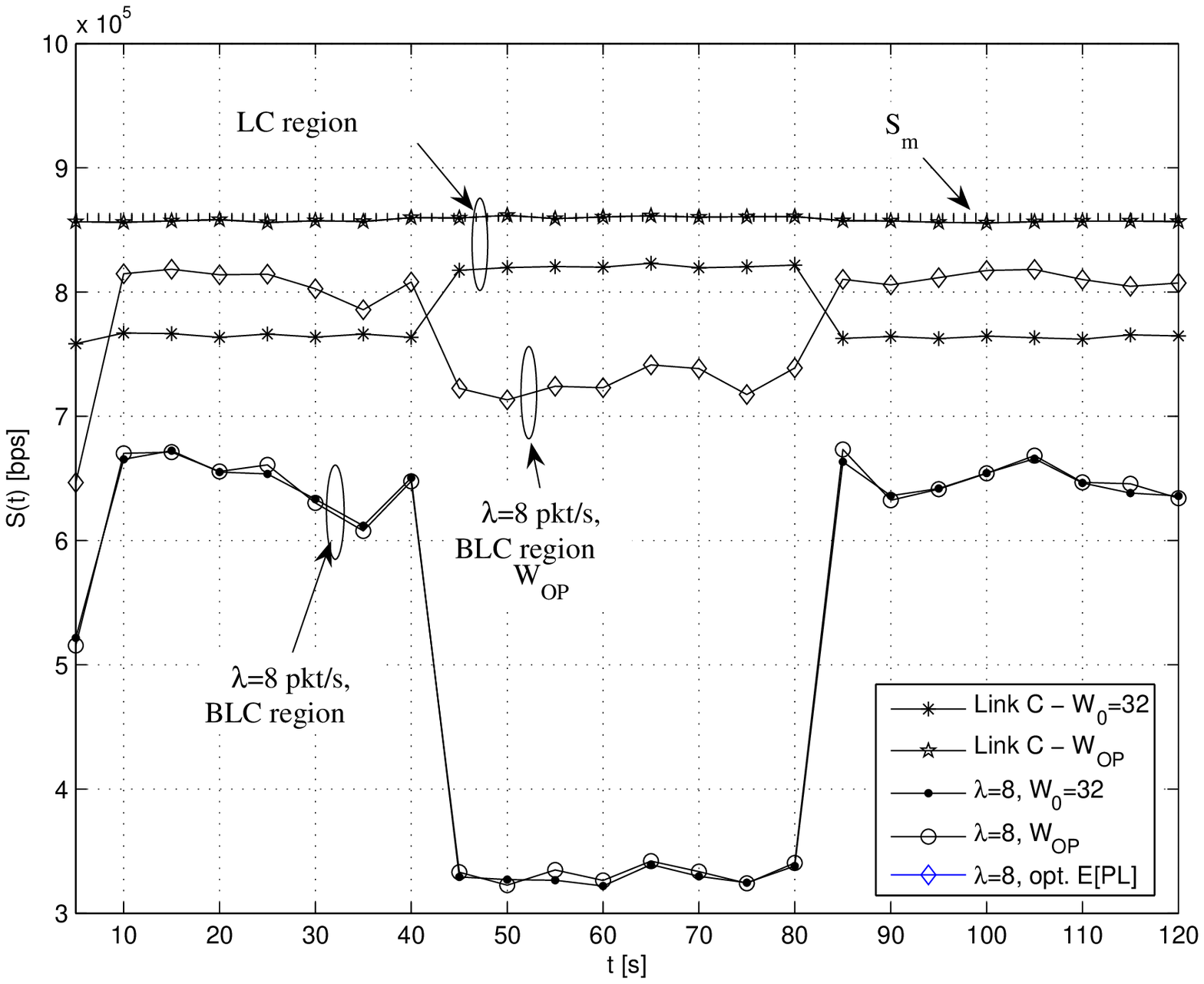}{Aggregate throughput of a network
with a maximum number of stations equal to 10, transmitting at the
fixed bit rate 1Mbps. The other network parameters are described
in the paper.}{final_simulation}
\clearpage
\begin{table}\caption{Typical network parameters}%\label{parameters}
\begin{center}
\begin{tabular}{c|c||c|c}\hline
\hline MAC header & 24 bytes & Slot time $\sigma$ & 20 $\mu s$\\
\hline PHY header & 16 bytes & SIFS & 10 $\mu s$\\
\hline ACK & 14 bytes & DIFS & 50 $\mu s$\\
\hline $\tau_p$ & $1\mu$s  &EIFS & 300 $\mu s$\\
\hline $W_0$ & 32&  ACK timeout & 300 $\mu s$\\

\hline m & 5 &CTS timeout & 300 $\mu s$\\

\hline \hline
\end{tabular}
 \label{tab.design.times}
\end{center}
\end{table}
%\hline Payload size & 1024 bytes\\
%\hline RTS & 20 bytes&EIFS & 300 $\mu s$\\
%\hline CTS & 14 bytes\\
%
%\begin{eqnarray}\label{straight_line_model}
%S(\lambda)\left|_{\lambda\approx 0^+}\right.&=&\frac{P_t \cdot
%P_s\cdot (1-P_e)E[PL]}{E[S_{ts}]}\nonumber\\
%&\approx& N E[PL]\frac{q}{E[S_{ts}]}\approx N\cdot
%E[PL]\cdot\lambda
%\end{eqnarray}
%
\clearpage
\begin{table}
  \centering\label{Optimization Algorithm}
  \caption{Optimization algorithm accomplished by every contending station.}
  \begin{tabular}{rl}
    %\hline
    %\multicolumn{2}{l}{Optimization Algorithm}\\
    \hline
    1. & \verb"do" \\
    2. & \qquad Estimation of PER and N  \\
    3. & \qquad Evaluate $\lambda_c$ from (\ref{lambda_critici})\\
    4. & \verb"while(IDLE)" \\
    5. & // Request to Send from Upper Layers:\\
    6. & \verb"if ("$\lambda > \lambda_c$\verb")" \\
    7. & \qquad // Link Capacity Operating Region: \\
    8. & \qquad The station evaluates $W_{opt}$ from (\ref{W_optimal}) \\
    9. & \verb"else" \\
    10. & \qquad // Below Link Capacity Operating Region:\\
    11. &\qquad compute $E[PL]^*_{1}$ given $\lambda_c$ \\
    12. & \qquad compute $E[PL]^*_{2}$ given $P_e$ \\
    13. &  \qquad select $E[PL]_{opt} = \min(E[PL]^*_{1},E[PL]^*_{2}, PL_{max})$ \\
    14. & \verb"end" \\
    15. & send packet \\
    \hline
  \end{tabular}
\end{table}
\end{document}